\documentclass[12pt]{article}

\newcommand{\RR}{{\rm R}}

\newcommand{\RRRR}{\RR^{4}}
\newcommand{\CC}{{\cal C}}

\newcommand{\polje}{\phi}
\newcommand{\fakt}{f}

\newcommand{\PQS}{the physics of quantum scattering}

\begin{document}

\ \vskip45mm
\parbox{130mm}{\Large\bfseries\centering Perturbative $S$-matrices that depend on parameters of a realistic regularization}
\vskip5mm
\parbox{130mm}{\centering Marijan Ribari\v c and Luka \v Su\v ster\v si\v c\footnotemark \\Jo\v zef Stefan Institute, p.p.3000, 1001 Ljubljana, Slovenia }

\footnotetext{Corresponding author. Phone +386 1 477 3258; fax +386 1 423 1569; electronic address: \tt luka.sustersic@ijs.si\rm}


\vskip 20mm

\begin{abstract}

We propose a partial answer to the question of what kind of ultrahigh-energy physics has to be taken into account to circumvent the appearance of ultraviolet divergencies; a more than sixty years old open question in quantum electrodynamics. To this end we introduce a new theoretical concept to the theory of quantum scattering: Perturbative $S$-matrices that depend on parameters of a realistic regularization---realistic in the sense of Pauli and Villars [Rev. Mod. Phys. \bf 21\rm, 434 (1949)]. We expect that these additional parameters may provide some new information about \PQS. There are such perturbative $S$-matrices also in the presence of non-renormalizable interaction terms with no counterterms. 

\end{abstract}
\vskip.5in

\it I. Introduction and motivations.---\rm Quantum field theory (QFT) provides us with a perturbative theory, whose physical content we can represent by the Feynman rules, cf. e.g. \cite{Bjorken, Hooft, Veltman}. Using these rules, we obtain perturbative expansions of the momentum-space $n$-point Green functions (Feynman series) to any desired order, written mainly in terms of ultraviolet-divergent integrals. Within the framework of QFT, Feynman series result, through various regularizations and renormalization, in the QFT perturbative $S$-matrices.

All regularizations used in QFT are formalistic \cite{Pauli} computational aids. Their parameters are eventually eliminated during renormalization by limiting them towards such values that the QFT perturbative $S$-matrices do not depend on the choice of regularization. 

We propose that, in computing the perturbative $S$-matrices from the regularized Feynman series, we retain the dependency on the regularization parameters, and thus try to gain some new information about \PQS. However, regularizations used in QFT involve either unphysical particles with negative metric or wrong statistic, or discrete space-time, or lowering the dimensionality of spacetime, or some combination thereof. Thus, it would be hard not to eliminate their regularization parameters and consider them as quantities related to \PQS.
 
In the Preface to \cite{Hooft}, 't Hooft and Veltman gave good arguments for taking the original, \it non-regularized \rm Feynman series as the most succinct representation of our present knowledge about \PQS. So we intend to use these series as a base for constructing perturbative $S$-matrices that depend on parameters of a realistic regularization \cite{Pauli}, which may provide some \it new information about \PQS. \rm

It is the primary purpose of this paper to propose, for the first time, that constructing a perturbative $S$-matrix without formalistic regularization, \it we should put aside \rm over a sixty-year old open question what kind of ultrahigh-energy physics has to be taken into account to circumvent the appearence of ultraviolet divergencies; instead we should concentrate on searching for the possible regularizing modifications of Feynman propagators that a theory about \it ultrahigh-energy physics could imply. \rm Using so modified Feynman propagators to regularize the Feynman series, we intend to obtain such a perturbative $S$-matrix that (i)~involves only the same particles as the QFT one, (ii)~depends on the additional parameters introduced by the modifications of Feynman propagators, (iii)~can be made equal to the QFT one, and (iv)~may provide some information about ultrahigh-energy physics.

\it II. Construction of perturbative $S$-matrices that depend on parameters of a realistic regularization.---\rm For a given non-regularized Feynman series and perturbative QFT $S$-matrix, let us consider construction of perturbative $S$-matrices that depend on  parameters of a realistic regularization.

As it seems that the vertices of non-regularized Feynman series adequately describe interactions in quantum scattering, it is taken that their ultraviolet divergencies are due to the asymptotic, high-energy behaviour of the Feynman propagators \cite{Villars}. So it is a prudent, conservative approach to retain the vertices in Feynman series, and modify only the Feynman propagators to create so regularized Feynman series that the Lehmann-Symanzik-Zimmermann reduction formula provides a perturbative $S$-matrix that: (i)~is Lorentz-invariant and unitary; (ii)~involves only the same particles as the QFT one; (iii)~depends solely on parameters that specify the QFT one and the modification of the Feynman propagators---for particular values of these parameters it is equal to the QFT perturbative $S$-matrix; and (iv)~exhibits the same symmetries as the QFT perturbative $S$-matrix.

Now, we can search for such modified propagators that effect a realistic regularization by considering the perturbative $S$-matrices they provide and their \it ultrahigh-energy behaviour; \rm in particular, we can evaluate the significance of their modifications on the basis of experimental data by statistical methods. We can facilitate this search by establishing the necessary and sufficient analytic properties of so modified Feynman propagators that they provide a perturbative $S$-matrix as specified above. As this is beyond the scope of this paper, we give only an example.

\it An example of sufficient analytic properties.\rm---In the case of QFT of a single real scalar field with $\polje^4$ self-interaction, we found out \cite{mi001,mi002} that it suffices that the modified Feynman propagator equals
\begin{equation}
   (k^2 + m^2 - i\epsilon )^{-1} \fakt(k^2 - i\epsilon) \,, 
       \qquad \epsilon \searrow 0 \,, \quad k \in \RRRR 
   \label{propagator}
\end{equation}
where the modifying factor $\fakt(z)$ that multiplies the QFT Feynman propagator has the following analytic properties: (a)~it is an analytic function of $z \in \CC$ except somewhere along the segment $z \le z_d < -9m^2$ of the real axis; (b)~$\fakt(-m^2) = 1$; (c)~$\fakt(z)$ is real for all real $z > z_d$, so that $\fakt(z^*) = \fakt^*(z)$; (d)~we can estimate that for all $z \in \CC$, 
\begin{equation}
   | \fakt(z) | \le a_0 ( 1 + | z |^{3/2} )^{-1} \,, \qquad a_0 > 0 \,;
   \label{faktest}
\end{equation}
and that for any real $z_0 > z_d $ the derivatives $\fakt^{(n)}(z)$ of $\fakt(z)$ are such that
\begin{equation}
   \sup_{ z \in \CC, \Re z \ge z_0} (1 + |z|^{3/2}) (1 + |z|^n) | \fakt^{(n)}(z) | < \infty \,, \quad n = 1, 2, \ldots ;
   \label{ocenaodvoda}
\end{equation}
(e)~the coefficients of $\fakt(z)$ depend on a positive cut-off parameter $\Lambda$ so that for any $\Lambda \ge \Lambda_0$, $\Lambda_0 > 0$, the modifying factor $\fakt(z)$ has properties (a)--(d) with the constants $z_d$ and $a_0$ independent of $\Lambda$, and we can estimate that:
\begin{equation}
   \sup_{\Lambda > \Lambda_0} \sup_{z > 0} | z^n \fakt^{(n)} (z)| < \infty \qquad
       \hbox{for} \quad n = 0,1,\ldots ,
   \label{regucond0}
\end{equation}
and that, in the asymptote of infinite cut-off parameter,
\begin{displaymath}
   \sup_{|z| < z_0} |\fakt^{(n)}(z) - \delta_{n0}| \to 0 \quad{\rm as} \quad 
        \Lambda \to \infty \quad\hbox{for any}\quad z_0 > 0 , 
                     \quad n = 0,1,\ldots. 
\end{displaymath}

One can verify that the functions
\begin{equation}
   (1 + \sqrt{ 1 - \mu m^2})^{n'} ( 1 + \sqrt{1 + \mu z } )^{-n'} , 
        \quad 0 < \mu < (3m)^{-2}, \quad n' = 3, 4, \ldots ,
   \label{primercek}
\end{equation}
may be used as a modifying factor, since they satisfy above conditions (a)--(e) with $\Lambda = \mu^{-1/2}$ and $\Lambda_0 > 3m$. As such modifying functions $\fakt(z) \to 1$ if $\mu \to 0$, the parameter $\mu$ determines the extent of modification of the Feynman propagators and of QFT $S$-matrix. So we can get a measure of the physical significance of such a modification for a particular process on the basis of the corresponding experimental data by determining the value and confidence intervals of $\mu$ through statistical methods.

\it Non-renormalizable interactions.\rm---If the interaction Lagrangian contains non-renormalizable terms that do not satisfy the Dyson criterion, e.g.~the Pauli term, that makes no difference to the above construction \cite{Gomis}. In contrast to QFT, \it no counterterms with free parameters are needed, \rm cf. \cite{Weinberg}. However, if the cut-off parameter (such as specified above) tends to infinity, the perturbative $S$-matrix might tend to infinity unless the coefficients of the non-renormalizable interaction terms vanish simultaneously.

\it III. Non-perturbative theory of quantum scattering.---\rm Eventually, there is an open question about the kind of non-perturbative theory we should look for incorporating such a perturbative $S$-matrix that depends on parameters of a realistic regularization. 

In QFT, the path-integral formalism provides the most direct way to Feynman series in their Lorentz-invariant form, especially so for non-Abelian gauge theories, cf. \cite{Weinberg, Kaku}. And according to Feynman, it also gives an insight into \PQS.

All this suggests we should try to figure out how to incorporate each of the above perturbative $S$-matrices into some path-integral based non-perturbative theory, which possibly takes account of gravitation, faster-than-light effects, and/or some ultrahigh-energy physics; for an example see \cite{mi002, mi003}. Which is all beyond the scope of this paper. 

\it IV. Conclusions.---\rm We propose construction of such perturbative $S$-matrices with possible non-renormalizable terms that might already provide some new information about \PQS\ on the basis of available experimental data.

\end{document}